\documentclass[twocolumn,aps]{revtex4}
\input{epsfig.sty}
\newcommand{\beq}{\begin{eqnarray}}
\newcommand{\eeq}{\end{eqnarray}}
\begin{document}
\title{Basic Treatment of QCD Phase Transition Bubble Nucleation}
\author{Leonard S. Kisslinger$^\dagger$ and Sameer 
Walawalkar$^\dagger$$^\dagger$\\
Department of Physics, Carnegie Mellon University, Pittsburgh, PA 15213\\
Mikkel B. Johnson$^\dagger$$^\dagger$$^\dagger$ \\
Los Alamos National Laboratory, Los Alamos, NM 87545}
\date{\today}
\begin{abstract}
Starting from the QCD Lagrangian and the surface tension of QCD bubbles
we derive the critical size of bubbles, the nucleation probability
and the nucleation site separation distance.

PACS Indices:12.38.Lg,12.38.Mh,98.80.Cq,98.80Hw
\vspace{3mm}

Keywords: Cosmology; QCD Phase Transition; Bubble Nucleation
\end{abstract}
\maketitle
\indent 

\section{Introduction}

  For more than two decades there have been studies of the nucleation of 
bubbles during the QCD phase transition (QCDPT), the transition from a 
universe of deconfined gluons, quarks and antiquarks (the quark/gluon plasma)
to a hadronic universe. These studies, based on 
a first-order QCDPT, use models of QCD to estimate the critical size of
bubbles, the nucleation rate, and the distance between nucleating 
centers\cite{h83,l83,kk86,fma88,ck92,ign,mr95,rmp95,cm96,nm,is01}. In an 
attempt to find possible observational effects from this early universe 
phase transition, there have also been estimates of magnetic fields 
generated by nucleating bubble surfaces during the QCDPT\cite{co,soj}, 
and possible observable CMBR effects arising from magnetic walls\cite{lsk03} 
which arise from collisions of nucleating bubbles\cite{jhk2003}. Estimates 
of magnetic fields arising from bubble collisions during the earlier 
electroweak phase transition (EWPT)using effective Lagrangians also have
been made\cite{kv,ae,cst}. Many of these studies make use of the Coleman 
model\cite{col} of tunneling from a false to a true vacuum. Nucleation
rates were estimated using standard thermodynamics/statistical 
mechanics\cite{fma88} and using hydrodynamics\cite{ck92a} with the
Langer formalism\cite{l69}.

   The knowledge that the universe evolved from a quark/gluon plasma to
our hadronic universe inspired the RHIC (relativistic heavy ion collisions)
program at BNL and other laboratories. The challenge for the RHIC program,
as well as the early universe studies, is to identify unambiguous observables
for the transition (or transitions). In a recent review\cite{uh04} possible
relations between the early universe and heavy ion heavy phase transitions
and the status of theoretical attempts to reach thermalization on the time 
scale that seems to be needed at RHIC are discussed. There is also a detailed
discussion of flow that is associated with RHIC. In this review,  as well as
in another recent report\cite{gm04}, the possible existence of color 
superconducting phases and a color glass condensate
that involves low momentum fields with long time scales compared to what
is considered to be natural time scales are examined. Extensive references 
to theoretical and experimental research in RHIC physics, and possible 
relationships to the early universe chiral phase transition, are given in
these reviews.

   Lattice gauge calculations indicate that there is no first order EWPT 
transition nor consistency with baryogenesis in the Standard EW Model.
These calculations also show 
that with supersymmetric fields there can be a first order transition
and consistency with baryogenesis during the EWPT . Lattice 
calculations are not yet able to prove definitely whether there is a first
order QCDPT. Because of the great importance of possible observational
effects of a first order QCDPT, with bubble nucleation, collisions, and
possible observable magnetic and other effects, in the present note we
assume that there is a first order chiral phase transition and
investigate QCD bubble nucleation.

   The most notable aspect of the present work is that we start from the basic 
QCD Lagrangian, rather than an effective model, and use previous research
on nonperturbative QCD condensates to carry out the calculations.

In Sec. II we derive the critical radius for nucleating bubbles, and in
Sec. III we estimate the site separation for  nucleation of bubbles during 
the QCDPT.

\section{Critical radius during the QCDPT}
  
   In this section we estimate the critical radius, $R_c$, for bubble
nucleation, starting from the basic QCD Lagrangian.
Physically, the critical radius is attained when force due
to the pressure difference inside and outside the bubble, a volume effect,
equals that of the surface tension, an area effect. In the following section
we derive the nucleation probability and site separation, which requires
quite different methods. First, an outline of the paper is given

\subsection{Outline of Paper}

The starting point of our work is the basic QCD Lagrangian density:

\subsubsection{QCD Lagrangian Density and basic Equations of Motion}
The QCD  Lagrangian density for massless quarks is
\beq
\label{QCD}
\ {\cal L}^{QCD}&=&-\frac{1}{2}Tr[G \cdot G]
+\bar{q}_f \gamma^\mu(i\partial_\mu+gA_\mu)q_f
\eeq
where
\beq
\label{G}
    G_{\mu\nu} & = & \partial_\mu A_\nu -  \partial_\nu A_\mu
  -i g [A_\mu,A_\nu] \\
    A_\mu & = & A_\mu^a \lambda^a/2 , \nonumber
\eeq
with $\lambda^a$ the eight SU(3) Gell-Mann matrices, $[\lambda^a,\lambda^b]
= 2if^{abc}\lambda^c$. Minimizing the action, one obtains the general
QCD equations of motion (EOM)
\beq
\label{eom1}
\partial_\mu\partial^\mu A^a_\nu -\partial_\nu \partial^\mu A^a_\mu 
+ g f^{abc}(2 A^b_\mu\partial^\mu A^c_\nu \nonumber \\
- A^b_\mu \partial_\nu A^{\mu c}- 
  \partial^\mu A^b_\mu A^c_\nu ) 
+ g^2 f^{abc} f^{cef}\nonumber \\
 A^b_\mu A^{\mu e}A^f_\nu +g\bar{q}_f \gamma_\nu \frac{\lambda^a}{2}q_f &=& 0
 \; .
\eeq
 
   As we shall see, the last term in the EOM, Eq.(\ref{eom1}), which gives the 
quark-gluon interaction, plays a crucial role in determining the critical
radius of nucleating bubbles during the QCD chiral phase transition.

\subsubsection{Method for Deriving the Critical Radius}
 
   As mentioned above, the force at a bubble wall from the pressure difference
on the two sides of the wall, which drives the nucleation, increases
for given pressure difference as one extra power of the bubble radius than the
force from the surface tension, which tends to shrink the bubble. For the
calculation of the surface tension, $\sigma$, and pressure difference, 
$\Delta p$, we start with the basic QCD Lagrangian. 
To determine the critical radius we
minimize the action with respect to the radius, as in the models of 
Refs.{\cite{cst,col,ck92a}). These calculation are done near the critical
temperature, and do not require finite temperature field theory. The 
approximations used are dropping the higher-dimensional gluonic terms in
calculating the surface tension and using only pions as hadrons for the
effect of the quark condensate forming at the critical temperature. These
approximations are discussed in the next subsections in this section. 

   For this investigation only the action in the vicinity of the bubble
wall at a temperature near the critical temperature is needed, and we use 
an extension of the instanton model to SU(3) symmetry in Minkowski space. 
For the more general study of nucleation during the QCD chiral phase transition
this model must be generalized to treat the interior and exterior of the bubbles.

\subsubsection{Method for determining the Nucleation Probability and Site
Separation}
   
   For the determination of the nucleation site separation, the main goal of
the present paper, standard classical statistical mechanics is used to
determine the rate at which nuclei of bubbles of critical radius form. 
For this the pressure difference between the quark/gluon and hadronic phases
at the critical temperature, as well as the surface tension and critical
bubble size, are needed, and are obtained from the QCD Lagrangian density
within our instanton-like model.

  For the final calculation of the site separation distance the temperature 
dependence of the pressure difference is needed. For this we use the standard 
finite temperature field theory, with a reduction from four-dimensional
space-time to three-dimensional space at fixed temperature by replacing real
time by imaginary 1/temperature.
 
\subsection{SU(3) Ansatz for the color gauge field}

    We use the Lorentz gauge and an SU(3) ansatz, in which we 
remove the color dependence of Eq(\ref{eom1}) by extending the 
color fields, $A^a_\mu$, to SU(3) matrices:
\beq
\label{su3}
  \partial_\mu A^a_\mu &=& 0 \nonumber \\
    A^a_\mu & \rightarrow & i\frac{\lambda^a}{2} W_\mu \; .
\eeq
Note that this ansatz cannot be used in the Lagrangian, Eqs.(\ref{QCD},
\ref{G}), to define the action and obtain the EOM Eq.(\ref{eom1}), but it is
rather a method to apply a reasonable symmetry on the color field in the EOM.
It can be compared to the instanton ansatz\cite{bev}, with the color fields
given in terms of the SU(2) quantities $\eta^{a}_{\mu\nu}$ \cite{hooft},
but our prescription defined in Minkowski space leads to Lorentz covariant 
equations needed for studying nucleation and other time-dependent processes.
We are guided in our choice of parameters by the instanton liquid
model reviewed in Ref\cite{shu98}.
In using this ansatz one operates on Eq.(\ref{eom1}) by the color SU(3)
matrix  $\lambda^a$, uses the matrix extension of the color field given by 
Eq.(\ref{su3}) and the properties of the SU(3) generators to obtain the 
EOM for the $W_\mu$
\beq
\label{eom2}
 && \partial_\mu\partial^\mu W_\nu -\frac{3}{2}g (2 W_\mu\partial^\mu W_\nu
- W_\mu \partial_\nu W^\mu) +\nonumber \\ 
 && \frac{9}{4}g^2 W_\mu W^\mu W_\nu -\chi^V <\bar{q}q> \bar{N} W_\nu = 0 \; ,
\eeq
where we make use of the study\cite{lsk99} of the vector vacuum
susceptibility defined by the three-point function,
\beq
\label{tpf}
<g\bar{q} \gamma_\nu\lambda^a/2 q> = \chi^V \bar{N} <\bar{q} q> A^a_\nu~,
\eeq
 in terms of the quark condensate,
$<\bar{q} q>$. The parameter $\chi^V$ was determined from the study of the
vector three-point function for a nucleon\cite{lsk99}, while the factor
$\bar{N}$, the number of hadrons in a typical hadronic volume $V$, is needed
for our finite T study and will be discussed below.
 
  To determine the critical radius one needs to balance the forces near the
surface of the bubble. The essential variable for determining the critical
radius is $s = \sqrt{x^\mu x_\mu}$, and we use the form
\beq
\label{wmu}
     W^\mu &=& x_\mu W(s) \; ,
\eeq
with the gauge condition 
\beq
\label{gc}
       W'(s) &=& -\frac{4}{s} W(s) \; .
\eeq
The resulting EOM for the function $W(s)$ is
\beq
\label{eom3}
      W'' + \frac{5}{s} W' -\frac{3}{2}g(W^2 +sW W')  \nonumber \\
 +\frac{9}{4} g^2s^2 W^3  -\chi^V \bar{N}<\bar{q}q> W &=& 0 \; .
\eeq

\subsection{Surface tension, pressure, and critical radius}

  Since the purpose of the present paper is to estimate the critical radius of
nucleating bubbles and the site separation, and we are not attempting to
study the general problem of bubble nucleation and collisions during the
QCDPT, we shall concentrate on the bubble wall and the regions in the QGP 
and HP close to the bubble walls. We neglect some of the hadronic structure
in the HP and quark/gluon structure in the QGP. This allows us to directly 
use the QCD Lagrangian for the nucleation properties in the early stages of
the QCDPT, which has not been done previously, but the method must be
extended for the complete treatment of the QCDPT.

  The instanton representation of the color gauge field is known to give a
satisfactory representation of mid-range nonperturbative QCD. Although
instantons cannot be defined between hadronic and quark/gluon vacua, we
shall use an instanton-like representation of the bubble wall, as in our
earlier work\cite{lsk03,jhk2003}. This will lead to a modification of the 
variables used in the expression of the color gauge fields, as explained 
below.

   Let us first estimate the surface tension, one of the two parameters 
needed to determine the critical radius.
The surface tension is obtained by integrating the energy density, 
$ T^{00}$, through the bubble wall,
\beq
\label{surface}
 \sigma^{inst} & = & \int ds  T^{00} \nonumber \\
              & = & \int ds [\frac{\partial {\cal L}}{\partial (\partial_t W)}
 \partial_t W -g^{00} {\cal L}] \\
\label{sigma}
  & \simeq & \int ds [6W^2 -\frac{9}{2}g(s^2W^3 +\frac{g}{8} s^4 W^4)]. 
\eeq
Only the dominant gluonic contribution in Eq.(\ref{sigma}) will be used in
estimating the surface tension.

In Ref\cite{lsk02} it was shown that an instanton model for the bubble wall
is consistent with the surface tension estimated in lattice QCD calculations.
Based on this we use an instanton-like form near the surface of the bubble
wall. Recognizing that $W_\mu W^\mu = 0$ at $s=s_0$, the position of the 
bubble wall in the instanton model, and is nonzero in the region of 
approximately $s = s_0 \pm \rho$, we modify the variables used in 
Eq.(\ref{wmu}); and for  $W_\mu W^\mu$ and $W(s)$ we assume the instanton-like
form
\beq
\label{W}
          W_\mu W^\mu &=& \bar{s}^2 W(\bar{s})^2 \nonumber \\        
         W(\bar{s}) &=& \frac{C_W}{(\bar{s}^2 + \rho^2)^2} \; ,
\eeq
which satisfies the gauge condition Eq.(\ref{gc}) for $\bar{s}>\rho$. 
Note that with the metric obtained with the continuation from Euclidean
space, in which instantons are derived, to Minkowski space, 
$\bar{s} = \sqrt{r^2-t^2} -s_0$, the most important region for $W_\mu
W^\mu$ is the interval $-\rho\le\bar{s}\le\rho$.
To check the validity of the this form we
solve the EOM Eq.(\ref{eom3}) for W(s), as shown in Fig.1.
\begin{figure}
\begin{center}
\epsfig{file=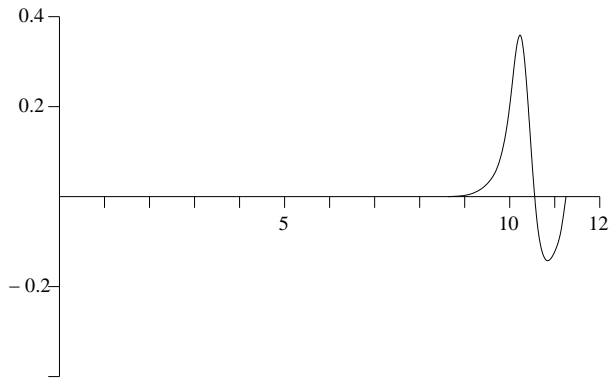,width=8cm}
\caption{Solution for W(s) for t=0 near the surface at $s_0=r\simeq R$=10.2 fm}
{\label{Fig.1}}
\end{center}
\end{figure}
In Fig.1 $s_0 = 10.2$ fm, which can be interpreted as R, the radius of the 
bubble if we take t=0 as the time shown in the figure.
One can see from the figure that the surface peak drops about a factor of
two as s-R goes from 0 to 0.15, as expected from the form of Eq.(\ref{W}).
Therefore, the instanton-like form is seen to be valid near the surface of
the bubble. For our calculation, we take $\rho=$ 0.2 fm, which is consistent
with the result of Fig. 1 and the value of the $\rho$ for the instanton
liquid model\cite{shu98}.
One can find the value of the parameter $C_W$ from such solutions, but
as is shown below this is not necessary in the present work, where we only
derive the critical radius and the nucleation site separation.
The oscillations in $W(s)$ found outside the wall would be important for 
collisions, which we do not treat in the present work. See Ref\cite{jhk2003}.

Recognizing that the lowest-dimensional term
is dominant, the surface tension is 
\beq
\label{sig}
       \sigma & \simeq & \int ds 6 W(s)^2 \;\;=\;\; \frac{15 \pi C_W^2}
{2^4 \rho^7}.
\eeq
To determine the pressure difference, we exploit the results of the Higgs
model\cite{cst,col}, noting that the expectation value of the
Higgs field is replaced in our work by the quark condensate, which also
goes from zero to a finite value during the phase transition.  In the Higgs
model, in the thin wall approximation\cite{cst,col},
\beq   
\label{dpcol1} 
   T^{00}(|\phi|=\eta) - T^{00}(|\phi|=0) =\epsilon \lambda \eta^4\,
\eeq 
which is the negative of the last term in the Lagrangian density used in
Ref\cite{cst}.
Note, however, that
\beq
\label{dpcol2} 
\epsilon \lambda \eta^4\ =\Delta p~~ {\rm (Higgs~Model)}.
\eeq
Thus, in the Higgs model, as well as for our Lagrangian, $\Delta p$ is given
by the difference in $T^{00}$ across the bubble wall. With this observation,
we are able to use all the familiar results of finite-temperature field
theory for obtaining the consequences of our Lagrangian density for
nucleation.

   Note that at this stage of the nucleation process we are working in the 
brief time interval between $t_c$ and $t_f$, the time following $t_c$ when 
the universe has reheated (see the following section on nucleation).
Since the gluonic condensate does not vanish rapidly with $T \simeq T_c$,
the difference in the gluonic free energy on the two sides of the 
bubble wall is negligible
during the first stages of nucleation of interest here. For similar
reasons we also neglect the free energy difference of the noninteracting 
fermionic terms within and outside of the bubble, which play an important
part of the QCDPT after the time $t_f$ when there is an equilibrium 
nucleation process from the quark/gluon to the hadronic universe.
For example, in a more general treatment of nucleation and collisions
for the QCDPT, the perturbative hadronic and quark kinetic energy terms
in the QCD Lagrangian given by $\bar{q}_f(i\gamma^\mu \partial_\mu)q_f$ must
be considered\cite{fma88,ck92}.

 Since the quark condensate is the parameter for the chiral QCD phase 
transition  being treated in the present work, as one can see from the 
expression for $T^{00}$, given in Eq.(\ref{surface}),
in our theory starting with the QCD Lagrangian the pressure difference is 
given approximately by the term in  $T^{00}$ that is the negative of
the quark-gluon interaction, ${\cal L}^{q(int)}=- <g \bar{q} \gamma_\mu 
(\lambda^a/2) q> A^{a \mu} $, 
the last term in the Lagrangian density (\ref{QCD}).  Having determined the
surface tension and $\Delta p$ from our Lagrangian, the critical radius can
be obtained by minimizing the four-dimensional action\cite{col}
\beq
\label{se}
   S_E &=& 2 \pi^2 \sigma R^3 + \frac{\pi^2}{2} \Delta p R^4 
\eeq 
with respect to the bubble radius, $R$.

   In applying this formalism to our model we use the following concepts:
\vspace{2mm}

\hspace{5mm} i) Our bubble wall is instanton-like, not instantons. This
model follows from the success in obtaining an instanton-like interior
wall after bubble collisions with QCD bubbles having such an instanton-like 
bubble wall\cite{jhk2003}.
\vspace{2mm}

\hspace{5mm} ii) The wall is very thin, so a thin-wall approximation is
justified. We just need the difference between $T^{00}$ inside and outside
the bubble wall.
\vspace{2mm}

\hspace{5mm} iii) Making use of the fact that the quark condensate 
vanishes for $T\geq T_c$ outside the bubble wall and that during the phase 
transition when $T\simeq T_c$ inside the bubble pions dominate the 
hadronic phase, we assume that all of the nonvanishing quark condensate is in 
the pions inside the bubble. Using Eq.(\ref{tpf}),
the quark-gluon interaction Lagrangian density inside the bubble is
\beq
\label{Lin}
  {\cal L}^{q(int)} &\simeq& -\chi^V n_\pi <\bar{q}q>A^a_\nu  A^{a \nu} \; ,
\eeq
while outside the bubble
\beq
\label{Lout}
       {\cal L}^{q(int)} &\simeq& 0 \; ,
\eeq
where the color field $A^a_\nu$ is not of the instanton-like form given by 
Eqs.(\ref{su3}, \ref{wmu},\ref{W}) except at the bubble wall. A more general 
theory of gluonic structure in the hadronic phase is needed to discuss 
the interior, but for the purposes of the present paper we only need the 
pressure difference on the two sides of our bubble wall.
\vspace{2mm}

\hspace{5mm} iv) From this we conclude that the difference in $T^{00}$ 
inside and outside the bubble wall is
\beq
\label{dT00}
    \Delta [ T^{00}] &\simeq& \frac{\chi^V}{2}n_\pi <\bar{q}q> <A_\mu A^\mu> \; ,
\eeq
with $<A_\mu A^\mu>$ the mean value of $A^2$ in the wall.
Using Eqs.(\ref{su3},\ref{wmu},\ref{W}), this gives for the pressure 
difference inside and outside the bubble surface in the thin
wall approximation
\beq
\label{free}
  \Delta p & \simeq & \frac{\chi^V}{2}n_\pi <\bar{q}q> \bar{s}^2 W(\bar{s})^2 
 \\
\label{dpinst}
           & \simeq & \frac{\chi^V}{2}n_\pi <\bar{q}q> \frac{C_W^2}{16}
\frac{1}{\rho^6}\; ,
\eeq
where we have taken $\bar{N}$ = $n_\pi$, the pion number, and have 
evaluated $A^{a \mu}$ at $\bar{s} = \rho$.  The fact that the pions 
dominate the hadronic density at the phase transition, and thus that the 
effective density of quark condensate is given by the pion density, is
 well-known.

To estimate the error in evaluating $\bar{s}^2 W(\bar{s})^2$ at $\bar{s}=\rho$,
we can calculate the mean value by calculating $I_a \equiv \int_{-a}^a 
\bar{s}^2 W(\bar{s})^2/2a$. The most reasonable value for $a$ is $a=\rho$.
One finds $I_\rho = \frac{C_W^2}{16}\frac{1}{\rho^6}\times 1.12$. This would
give a 12$\%$ decrease in the critical radius and a larger nucleation site
separation. If one takes $a=2 \rho$, which is larger than expected for the
instanton-like wall, $I_\rho = \frac{C_W^2}{16}\frac{1}{\rho^6}\times 0.82$.
Therefore we conclude that the estimate in Eq.(\ref{dpinst}) is correct to
about $10\%$.

  Recognizing that the bubble wall thickness is given by $\rho = 0.2fm$, we
use the thin-wall approximation in minimizing the action\cite{cst,ck92a} 
to get the classical thin-wall equation for the critical radius.
From Eqs.(\ref{sig},\ref{dpinst},\ref{se}) we obtain
\beq
\label{rc}
  R_c &=& \frac{3 \sigma}{\Delta p} \;\; \simeq \;\; \frac{15 \pi}{\rho} 
 \frac{1}{\chi^V <\bar{q}q>/2}\frac{1}{n_\pi} \; .
\eeq
Using the standard value for the pion volume density\cite{kap}, $n_\pi/V = 
0.365 T^3$, with $V = (4\pi/3)R_n^3$ = the nucleon volume ($R_n$=1.1 fm),
and using the value of the vector susceptibility from Ref\cite{lsk99}, \\
$\chi^V<\bar{q}q>/2 \simeq 3 GeV^2$, with $T=150 MeV$,
we obtain from Eq.(\ref{rc})
\beq
\label{rcresult}
        R_c &=& 11 fm \; .
\eeq

  There are a number of approximations which could change our value for
$R_c$, and due to the sensitivity of the nucleation site separation to
this parameter, this could change the evaluation in the next section. We
show that this value is consistent with the evaluation of 
Ref\cite{ck92} at the critical time of freezeout in the next section. The
innovation of the present work is that we estimate $R_c$ directly from the 
QCD Larangian, rather than using models as in previous estimates. 

\section{Nucleation probability and nucleation site separation}

   For applications to astrophysical observations the distance between
nucleation sites, $d_n$, is a critical parameter, since with large separation
between nucleation sites the hadronic universe can be formed via the
collision between a few large nucleating bubbles, and this could lead to
interesting large-scale structure. E.g., it was shown with such a scenario 
that large magnetic walls would form during the QCDPT\cite{lsk03}, which
could lead to observable effects in Cosmic Microwave Background polarization.

   One obtains the nucleation site separation, $d_n$, from the probability of
nucleation per volume-time, $p(t)$.  In this section we estimate the 
nucleation probability using the critical radius derived in the previous 
section. In classical statistical theory the rate at which nuclei form per 
unit volume-time is given by 
\beq
\label{prob}
    P(T(t)) &=& P_0 e^{-S(t)} \; ,
\eeq
where in a Coleman-type model of tunneling from the false to the true vacuum
the action, $S(t)$, is treated in Euclidean space, and for a nucleus of 
critical radius in four dimensions is
$S= 2\pi^2 \sigma R_c^3 -\frac{\pi^2}{2} \Delta p R_c^4$, the extremum of 
which gives the relationship of Eq.(\ref{rc}). Most of the work that has
been done during the past two decades has assumed homogeneous nucleation,
for which the form
\beq
\label{prob1}
    P(t) &=& P(t_f) e^{-\alpha(t_f -t)} \; ,
\eeq
with $t_f$ the time at which the universe has reheated after the time
$t_c$, at which time $T$ is the critical temperature $T_c$ for the phase
transition (assumed first order). The parameters of the theory are determined
from the energy momentum tensor in a field theory or from the thermodynamic 
potential in a theory using classical statistical mechanics. In the following
subsection we discuss the application of our QCD approach assuming homogeneous
nucleation, and in the next subsection briefly consider inhomogeneous
nucleation.

\subsection{Homogeneous nucleation}

   From statistical mechanics one knows\cite{fma88} that the probability
for a fluctuation producing a bubble of radius $R_c$ for $T<T_c$ satisfies
\beq
\label{PT}
   P(T) &\propto&  e^{-\Delta F/T} \; ,
\eeq
with $\Delta F$ the bulk free energy difference between the two phases plus 
the contribution from the surface free energy,
\beq
\label{DF}
     \Delta F &=& \frac{4\pi}{3} R_c^3 (P_q -P_h) + 4\pi \sigma R_c^2 \; ,
\eeq
with ($P_q,P_h$) the pressure in the (quark,hadronic) phases, and in our
previous notation $\Delta p = P_q -P_h$.
In the estimate of nucleation in the bag model\cite{fma88,ck92} the
quark pressure at $T<T_c$, with $B$ the bag constant, is
\beq
\label{pq}
    P_q &=& \frac{37 \pi^2}{90} T^4 - B \; .
\eeq 
The main difference between our present work and that of the earlier work
is that we use the QCD Lagrangian directly and the vanishing of the
quark condensate at $T=T_c$, rather than the bag model to obtain $\Delta F$.

   In the approach of \cite{ck92} Csernai and Kapusta, who use the Langer 
model\cite{l69}, and take $P_h = (3\pi^2/90) T^4$,
find a singularity in $R_c(T)$ as $T=T_c$, and obtain a value of $R_c$ in
agreement with our value of $R_c \simeq 11 \; fm$ for $(T_c-T)/T_c \simeq 1\%$.
Noting that  $(T_c-T_f)/T_c$ is of the order of $1\%$, one sees from 
Eq.(\ref{prob1}) that our
result for the critical radius $R_c$ is not inconsistent with this work.

   We now continue with the standard statistical mechanics 
approach\cite{fma88}.
To obtain the nucleation site number density one carries out the integral
\beq
\label{Nngen}
       N_n &=& \int_{t_c}^\infty p(T(t)) f(t) dt \nonumber \\
           & \simeq & \int_{t_c}^{t_f} p(T(t)) \; ,
\eeq
where $f(t)$ is the fraction of the universe which has not nucleated at time
$t$, and Eq(\ref{Nngen}) uses the result of Ref\cite{kk86} that $f(t)$ is
a step function quickly disappearing at the time $t_f$ shortly after $t_c$.
In this picture one obtains for the site separation distance\cite{fma88}
\beq
\label{dn}
   d_n &\simeq& 0.3 \frac{\sigma^{3/2} t_c}{T_c^{1/2} L} \; .
\eeq
The latent heat density is obtained from the free energy difference
between the phases by\cite{fma88}
\beq
\label{latent}
    L &=& T_c \frac{\partial}{\partial T} \Delta p \; .
\eeq
Since the $T$ dependence of $\Delta p$ is needed, one cannot use the result
of Eq(\ref{dpinst}), but must return to the thin-wall relationship between
$\Delta p$ and  $W^\mu W_\mu$, Eq(\ref{free}). 
Using our observation that in the vicinity of the bubble walls the solution
for $W^\mu W_\mu$ has the instanton-like form of Eq(\ref{W}), we find that
\beq
\label{dpl}
        \Delta p &=& K \frac{\bar{s}_3^2}{(\bar{s}_3^2 +\rho^2)^2} \; ,
\eeq
with K a constant. In real time Euclidean space $\bar{s}_3^2 =(r-R)^2 +t^2$. 
Making the reduction from 4- Euclidean space to equilibrium, 
$t^2 \rightarrow -1/T^2$, and $\bar{s}_3^2 = (r-R)^2 -(1/T)^2$. Using this form 
and the fact that for $\bar{s}_3^2 = \rho^2$, $K=4 \rho^2 \Delta F=
\frac{3\sigma}{R_c}$, we find that
\beq
\label{L}
     L &=& \frac{4}{T_c^2 \rho^2}\frac{3 \sigma}{R_c} \; .
\eeq
Using the values $\rho = .2 fm$, the lattice gauge value for $\sigma$,
and our result that $R_c = 11.0 fm$, we find from Eq(\ref{dn})
\beq
\label{dnour}
   d_n &=& 5.23 m \; .
\eeq
As anticipated, this value is about an order of magnitude larger than
previous bag model results.

\subsection{Inhomogeneous nucleation}

   The idea that impurities can have an important effect on the probability
of nucleation for cosmological phase transitions has been known for over
two decades\cite{ho83}. Recently this has been considered for the 
QCDPT\cite{cm96,is01}. As pointed out in \cite{cm96}, impurities in the
universe, with number density $n_{in}$ at a time $t_i,\; t_c<t_i<t_f$, can be
represented by a $P(T(t))$ which differs from the homogeneous value of 
Eq.(\ref{prob1}) by
\beq
\label{probin}
  P(t) &=& n_{in}\delta(t-t_i) +  P(t_f) e^{-\alpha(t_f -t)} \; .
\eeq
Using models for this form the nucleation site separation can be orders
of magnitude larger than the value for homogeneous nucleation.

   We would of course obtain similar model-dependent results, but since
this does not follow directly from our QCD picture we do not consider it 
further here.

\section{Conclusions}

   In this work we have started from the energy-momentum tensor derived from
the basic QCD Lagrangian with quark and gluonic color fields and assumed a
first-order QCD phase transition when the temperature of the universe is
$T_c \simeq 150 MeV$. From this $T^{00}$ the surface tension of gluonic 
walls in an instanton-like SU(3) treatment of the color field is derived. 
The free 
energy difference between the two phases on the opposite sides of the bubble
wall arises mainly from the vanishing of the quark chiral condensate in
the quark/gluon phase, and we derive this $\Delta F$ using previous work
on the nonperturbative vector quark three-point function. Recognizing that 
the wall is only 0.2 fm thick, the classical action is minimized in the
thin-wall approximation to derive the critical radius for nucleation, $R_c$,
for which the pressure difference driving nucleation overcomes the surface
tension contracting the bubble. We find that $R_c \simeq 11\; \; fm$, which
is satisfactory for possible astrophysical observations following from
the QCDPT nucleation and bubble collisions. Assuming standard homogeneous
nucleation, using our results for the critical radius, and deriving
the latent heat difference density from the instanton model of QCD and the
quark condensate free energy difference, we find that the separation
of nucleation sites is of the order of several meters, more than an order of 
magnitude larger than the standard QCD Bag model that has been used by 
previous authors. As research by other investigators has shown, the 
possibility of inhomogeneous nucleation could lead to quite large distances 
between nucleation sites and a QCDPT with only a few active bubbles. In our 
next research on this topic we shall investigate possible observable 
magnetic structures that would arise from QCD nucleation processes.
\vspace{.5 cm}

\noindent {\bf Acknowledgements}
\vspace{.5 cm}

  This work was supported in part by the NSF grant PHY-00070888, in part 
by the DOE contract W-7405-ENG-36. The authors thank Profs. Ho-Meoyng Choi,
Ernest Henley and W.Y. Pauchy Hwang for helpful discussions.
\vspace{.5cm}

\noindent
$\dagger$ email:kissling@andrew.cmu.edu\\
$\dagger$$\dagger$ email:sameer@andrew.cmu.edu\\
$\dagger$$\dagger$$\dagger$ email:mbjohnson@lanl.gov


\end{document}